\documentclass[webpdf,contemporary,large]{oup-authoring-template}%

\graphicspath{{Fig/}}

\usepackage{natbib}

\theoremstyle{thmstyleone}%
\theoremstyle{thmstyletwo}%
\theoremstyle{thmstylethree}%

\expandafter\def\expandafter\normalsize\expandafter{%
    \normalsize%
    \setlength\abovedisplayskip{0pt}%
    \setlength\belowdisplayskip{8pt}%
    \setlength\abovedisplayshortskip{-6pt}%
    \setlength\belowdisplayshortskip{2pt}%
}

\begin{document}

\journaltitle{Bioinformatics}
\DOI{DOI HERE}
\copyrightyear{2025}
\pubyear{2025}
\access{Advance Access Publication Date: Day Month Year}
\appnotes{Application Note}

\firstpage{1}


\title[\texttt{pared}: Model selection using multi-objective optimization]{\texttt{pared}: Model selection using multi-objective optimization}

\author[1,$\ast$]{Priyam Das\ORCID{0000-0003-2384-0486}}
\author[2]{Sarah Robinson}
\author[3]{Christine B.\ Peterson\ORCID{0000-0003-3316-0468}}

\authormark{Das et al.}

\address[1]{\orgdiv{Department of Biostatistics}, \orgname{Virgina Commonwealth University}}
\address[2]{\orgdiv{Department of Statistics}, \orgname{Rice University}}
\address[3]{\orgdiv{Department of Biostatistics}, \orgname{The University of Texas MD Anderson Cancer Center}}

\corresp[$\ast$]{Corresponding author. \href{dasp4@vcu.edu}{dasp4@vcu.edu}}

\received{Date}{0}{2025}
\revised{Date}{0}{2025}
\accepted{Date}{0}{2025}



\abstract{\textbf{Motivation:} Model selection is a ubiquitous challenge in statistics. For penalized models, model selection typically entails tuning hyperparameters to maximize a measure of fit or minimize out-of-sample prediction error. However, these criteria fail to reflect other desirable characteristics, such as model sparsity, interpretability, or smoothness. \\ \vskip-.3cm
\textbf{Results:} We present the R package \texttt{pared} to enable the use of multi-objective optimization for model selection.
Our approach entails the use of Gaussian process-based optimization to efficiently identify solutions that represent desirable trade-offs.
Our implementation includes popular models with multiple objectives including the elastic net, fused lasso, fused graphical lasso, and group graphical lasso. Our R package generates interactive graphics that
allow the user to identify hyperparameter values that result in fitted models which lie on the Pareto frontier. \\ \vskip-.3cm
\textbf{Availability:} We provide the R package \texttt{pared} and vignettes illustrating its application  to both simulated and real data at \url{https://github.com/priyamdas2/pared}. \\ 
 }

\maketitle

\section{Introduction} \label{sec:intro}
Model selection is a classic and well-studied statistical problem. It is a non-trivial challenge to select a model that balances the trade-off of model parsimony, fit, and generalizability. To paint a picture of this challenge in the high-dimensional setting, we begin by considering the lasso \citep{lasso}, as a classic penalized regression approach with a single penalty parameter. The choice of this penalty parameter, typically denoted $\lambda$, is critical, as the value of $\lambda$ determines both the estimated coefficient values as well as the model sparsity.
Typically, $\lambda$ is chosen using cross-validation, which utilizes data-splitting to estimate prediction error on a held-out test set \citep{wu2020survey}. Given a grid  of $\lambda$ options, the optimal value is selected as the one that minimizes the out-of-sample prediction error. 

Although simple, parameter selection via cross-validation has several key drawbacks. Firstly, it can lead to instability, as different random splits of the data will lead to different selected values of the parameters \citep{roberts2014stabilizing}. Secondly, in practice, it may result in models that include more features than desirable, as the optimization is focused around prediction accuracy rather than sparsity. Finally, it may be time-consuming to re-fit the model for all the values on the grid. This may be particularly inefficient if the grid contains many values that are far away from the optimum. As an additional challenge, prediction error on a test set is not a sensible target for unsupervised learning problems such as clustering or graphical model inference
 \citep{Li2013}. 

As an alternative, fit criteria may be applied for model selection. Classical measures of fit include the information criteria, most notably AIC \citep{AIC} and BIC \citep{BIC}. These may be applied to penalized regression models, but lack theoretical support for the $p > n$ setting and tend to select overly dense models \citep{Zou2007}.  
 Another vein of research relies on stability under resampling as a criteria for model selection \citep{stability_selection, stars}. However, this approach inherits limitations  of the base learner: for example, the lasso tends to select only one from among a set of correlated features. This may result in low stability for correlated predictors with lasso as the base model. 

The problem becomes even more challenging in models that require the choice of multiple hyperparameters. In particular, regression models that seek to address limitations of the lasso may incorporate more complex penalty structures with multiple tuning parameters. For example, 
 the elastic net \citep{Zou2007} and fused lasso \citep{flasso}, which target smoothness or grouping of the coefficients in addition to sparsity, require the choice of two tuning parameters. In the graphical modeling framework, the graphical lasso relies on a single $\ell_1$ penalty to achieve sparsity in the precision matrix \citep{friedman2008sparse}. Methods designed for the inference of multiple networks, including the fused and group graphical lasso \citep{jgl}, incorporate an additional penalty term to encourage sharing of common edges or similar edge values across multiple networks.

Critically, these methods seek to achieve multiple desired objectives, including sparsity and some version of smoothness or similarity.  Relying on a single measure of model fit for model selection neglects these desiderata. Pareto optimization, also known as multi-objective optimization, seeks to identify a set of solutions that lie along the Pareto front. These points represent optimal trade-offs in the sense that improving any of the criteria will necessarily worsen another of the targets. 

Here, we introduce the \texttt{pared} R package for \textbf{Pare}to-optimal \textbf{d}ecision making in model selection. 
The name \texttt{pared} implies that our approach enables the user to pare or trim away model complexity to reach a solution that represents a reasonable trade-off of fit, sparsity, and smoothness.

\section{Methods}\label{sec2}
Our proposed method is particularly suitable for models with multiple hyperparameters and multiple objectives.

\subsection{Statistical models}

We provide an implementation of our proposed approach for popular models including the elastic net, fused lasso, fused graphical lasso, and group graphical lasso.
Here, we briefly describe these methods, their hyperparameters, and the competing objectives that we aim to minimize.\\ \vskip-.2cm

\noindent \textbf{Elastic net}. 
The elastic net \citep{enet} combines $\ell_1$ and $\ell_2$ penalties to allow for both selection and regularization of the model coefficients. One attractive property of the elastic net is that  correlated predictors are likely to be selected together, whereas the lasso tends to arbitrarily include one predictor from among a correlated set.
The elastic net can be written:\\
\begin{equation*}
\hat{\beta} \equiv \underset{\beta}{\mathrm{argmin}}\bigg( \frac{1}{2n} \| y-X\beta\|^2 + \lambda\Big[ \alpha \|\beta\|_1 + (1-\alpha) \|\beta\|^2 \Big] \bigg).
\end{equation*}
Varying values of $\alpha$ capture the spectrum of models from lasso ($\alpha = 0$) to ridge ($\alpha=1$), while $\lambda$ controls overall regularization.

The two elastic net hyperparameters are typically selected using cross-validation on a grid. 
However, this approach has several drawbacks as outlined in Section \ref{sec:intro}.
In addition, 
recent work on elastic net points out that the marginal likelihood has a flat optimal region  across many $(\alpha, \lambda)$ combinations, which means that additional criteria besides fit need to be considered in model selection \citep{van2023fast}. 
In the \texttt{pared} R package, we consider the following objectives:  \vspace*{-.1cm}
\begin{itemize}
\item[{\tiny{$\bullet$}}] Fit: deviance 
\item[{\tiny{$\bullet$}}] Sparsity: number of non-zero coefficients 
\item[{\tiny{$\bullet$}}]  Shrinkage: $\ell_2$ norm of the coefficients 
\end{itemize}
Here, we consider increased shrinkage (smaller $\|\beta \|^2$) as a metric that reflects a reduced risk of overfitting. \\ \vskip-.2cm

\noindent \textbf{Fused lasso}. The fused lasso \citep{flasso} includes an $\ell_1$ penalty to induce sparsity and a penalty on the successive differences of the model coefficients to encourage smoothness across ordered variables.
The fused lasso model is as follows: \\
\begin{equation*}
\hat{\beta} \equiv \underset{\beta}{\mathrm{argmin}}\bigg( \frac{1}{2n} \| y-X\beta\|^2  + \lambda_1 \|\beta\|_1  + \lambda_2 \sum_{j=2}^p |\beta_j - \beta_{j-1} | \bigg).
\end{equation*}
The hyperparameters $\lambda_1$ and $\lambda_2$ control overall sparsity and sparsity in differences, respectively.
We consider the following objectives in model selection:  \vspace*{-.1cm}
\begin{itemize}
\item[{\tiny{$\bullet$}}] Fit: residual sum of squares
\item[{\tiny{$\bullet$}}] Sparsity: number of non-zero coefficients 
\item[{\tiny{$\bullet$}}]  Roughness: mean absolute difference of consecutive betas
\end{itemize}

\noindent \textbf{Joint graphical lasso}. The joint graphical lasso (JGL) \citep{jgl} seeks to estimates multiple graphical models, achieving overall sparsity through an $\ell_1$ penalty on the precision matrix for each group  $\mathbf{\Theta}^{(k)}$ and encouraging sharing across groups through a penalty on cross-group differences. The general form of the JGL is: \\
\begin{equation*}
\{\hat{\mathbf{\Theta}}\} \equiv \underset{\{\hat{\mathbf{\Theta}}\}}{\mathrm{argmax}} \bigg( 
\sum_{k=1}^K n_k \big[ \log \det ( \mathbf{\Theta}^{(k)})  - \text{tr}( \mathbf{S}^{(k)} \mathbf{\Theta}^{(k)} )  \big] - P(\{\hat{\mathbf{\Theta}}\})
\bigg),
\end{equation*} where $K$ is the number of sample groups, $n_k$ is the sample size for the $k$th group, $\mathbf{S}^{(k)}$ is the empirical covariance matrix for the $k$th group, and $P$ is a penalty function.

There are two variants of this model: the \textit{fused} and the \textit{group} graphical lasso, defined by the choice of penalty function $P$. The fused graphical lasso combines a within-group $\ell_1$ penalty with a fused lasso penalty designed to encourage similar edge values across groups:
\begin{equation*}
P(\{{\mathbf{\Theta}}\}) =  
\lambda_1 \sum_{k=1}^K\sum_{i\neq j} |\theta_{ij}^{(k)} |  +
\lambda_2 \sum_{k<k'} \sum_{i,j} |\theta_{ij}^{(k)} - \theta_{ij}^{(k')}|.
\end{equation*}
The group graphical lasso combines a within-group $\ell_1$ penalty with a group lasso penalty designed to encourage overlapping edge selection across groups:
\begin{equation*}
P(\{{\mathbf{\Theta}}\}) =  
\lambda_1 \sum_{k=1}^K\sum_{i\neq j} |\theta_{ij}^{(k)} |  +
\lambda_2 \sum_{i \neq j} \Big( \sum_{k=1}^K (\theta_{ij}^{(k)})^2 \Big)^{1/2}.
\end{equation*}
In both the fused and group graphical lasso, $\lambda_1$ controls model sparsity while $\lambda_2$ drives cross-group similarity.

For the fused graphical lasso, we consider the following three objectives: \vspace*{-.1cm}
\begin{itemize}
\item Fit: Akaike information criteria (AIC)
\item Sparsity: total number of selected edges across groups
\item Cross-group similarity: mean absolute difference of precision matrices across groups
\end{itemize}
For the group graphical lasso, we consider an overlapping set of criteria:  \vspace*{-.1cm}
\begin{itemize}
\item Fit: Akaike information criteria (AIC)
\item Sparsity: total number of selected edges across groups
\item Cross-group similarity: number of shared edges 
\end{itemize}

 \vspace*{-.1cm}
Our R package enables users to visualize trade-offs across these metrics and identify parameter combinations that yield desirable compromises using interactive graphics. In the accompanying GitHub vignette {\url{https://github.com/priyamdas2/pared}, we provide example use cases for the models described above. 

\subsection{Optimization approach}
Various algorithms have been proposed for hyperparameter optimization in the machine learning literature, including 
Bayesian optimization \citep{Snoek2012},
evolutionary algorithms \citep{karl2023multi}, and particle swarms \citep{xue2012particle}. Here, we seek to bridge this literature to provide an easy-to-use model selection tool for popular regression and graphical models in R. To do so, we rely on the R package \texttt{GPareto}, which provides tools for solving multi-objective optimization 
problems in settings where evaluating the objective functions is computationally expensive. This optimization framework uses Gaussian process (GP) models to emulate each objective and implements sequential optimization strategies to efficiently explore trade-offs between conflicting objectives. More precisely, the package addresses 
models with multiple outputs, $y^{(1)}(x), \ldots, y^{(q)}(x)$, where each $y^{(i)}: \mathcal{X} \subset \mathbb{R}^d \rightarrow \mathbb{R}$ is simultaneously optimized 
over the domain $\mathcal{X}$.

Because objectives are often conflicting (e.g., quality versus cost), no single solution minimizes all objectives at once. Instead, the goal is to approximate the Pareto set—the set of non-dominated solutions where no objective can be improved without worsening at least one other \citep{collette2003multiobjective}. The image of this set in the objective space forms the Pareto front, which helps users identify well-balanced trade-offs. Our proposed R package, \texttt{pared}, enables users to construct such Pareto fronts by balancing context-specific objectives beyond a single criterion, for popular statistical models designed to address conflicting objectives. 
Computation times are provided in Table~S1 of the Supplementary Material.
\begin{figure*}[htbp]
  \centering
  \includegraphics[width=\textwidth]{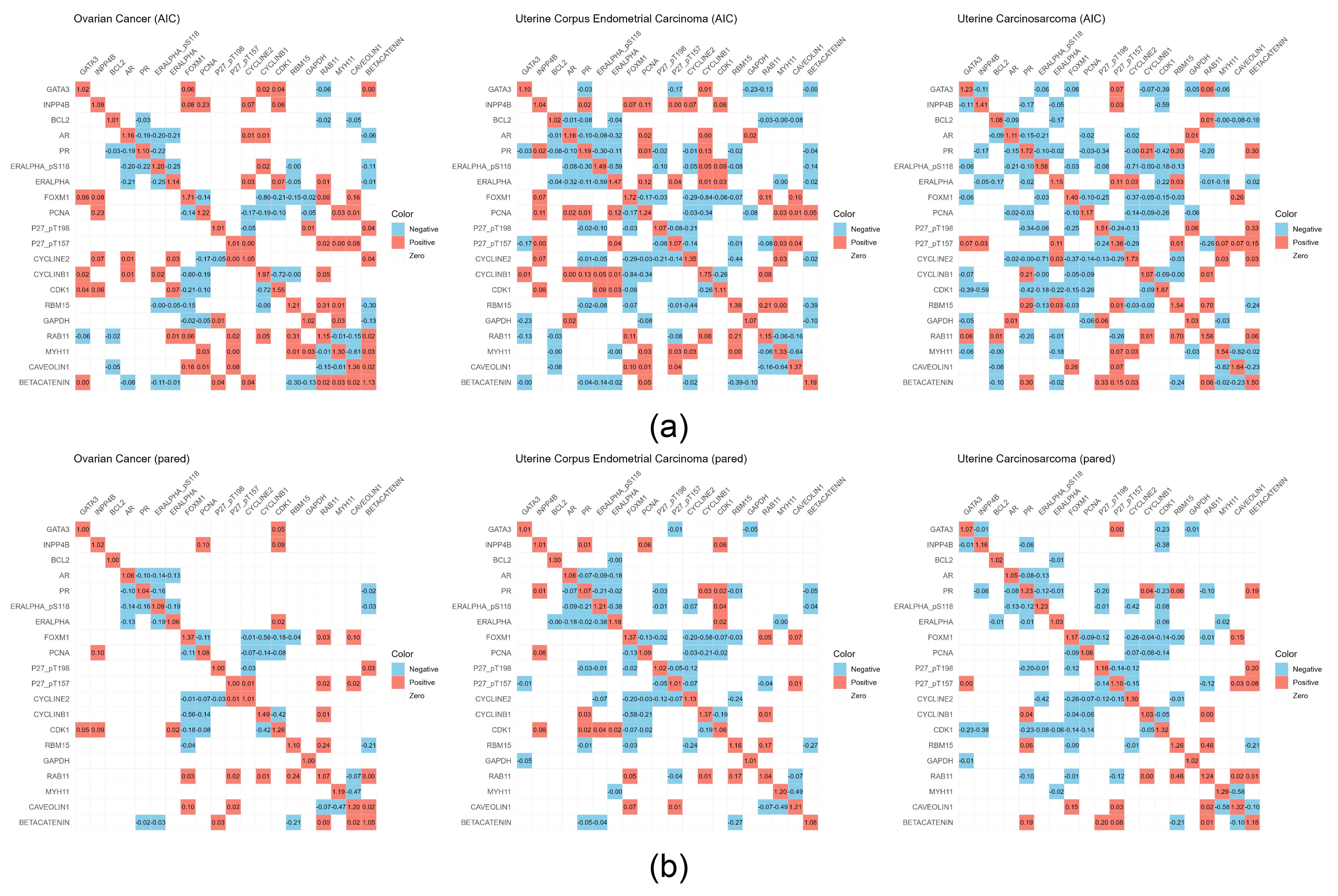}
  \caption{Estimated precision matrices obtained using the group graphical lasso for ovarian (OV), uterine corpus endometrial carcinoma (UCEC), and uterine carcinosarcoma (UCS) for the \textit{breast reactive}, \textit{cell cycle}, \textit{hormone receptor}, and \textit{hormone signaling breast} pathways. (a) shows the network corresponding to the model with minimum AIC. (b) displays a sparser network obtained using \texttt{pared}, representing one of the Pareto-optimal solutions.}
  \label{Precision_mat_comparison}
\end{figure*}

\section{Illustrative analysis}\label{sec3}
In this section, we illustrate the utility of our approach in selecting a parsimonious and smooth set of protein-signaling networks across gynecological cancer types.
To conduct this analysis, we obtained normalized protein abundance data from 
The Cancer Proteome Atlas \citep[TCPA, ][]{Li2013TCPA}, which 
quantifies protein markers using 
 antibodies targeting key oncogenic and cellular signaling pathways. Recent studies have highlighted that similar pathway activities are observed across the gynecological cancer types ovarian (OV), uterine corpus endometrial carcinoma (UCEC), and uterine carcinosarcoma (UCS) \citep{Berger2018}. Specifically, the pathways \textit{breast reactive}, \textit{cell cycle}, \textit{hormone receptor}, and \textit{hormone signaling breast} were found to be substantially activated in these cancer types \citep{Das2020}. 
 To leverage the potential structural similarity of signaling networks across these cancer types, we applied the group graphical lasso to simultaneously estimate the proteomic networks of OV, UCEC, and UCS using a set of 20 proteins associated with the aforementioned pathways, with sample sizes of 428, 404, and 48, respectively. To illustrate the standard model selection approach, we first applied the AIC, as recommended by \cite{jgl}. The corresponding estimated precision matrices characterizing the proteomic networks are shown in Figure~\ref{Precision_mat_comparison}(a).

Subsequently, we derived the Pareto-optimal set of models that balance AIC, sparsity, and the number of shared edges across networks using \texttt{pared}, with a run time of 2.4 minutes.  Figure~\ref{pared_interactive_plot} presents a screenshot of the interactive Pareto-front plot generated by \texttt{pared}, displaying the optimal set of solution points obtained using the group graphical lasso. Hovering the mouse cursor over any point reveals the corresponding model details, including tuning parameter values and model selection metrics. 
Figure~\ref{Precision_mat_comparison}(b) illustrates one such Pareto-optimal model, which is sparser than the model selected solely based on AIC. This demonstrates the potential of our approach to identify alternative optimal models that are more sparse, offering greater potential for network reliability and interpretability. 

\newpage

\begin{figure}[t!]
\centering
  \includegraphics[width=0.42\textwidth]{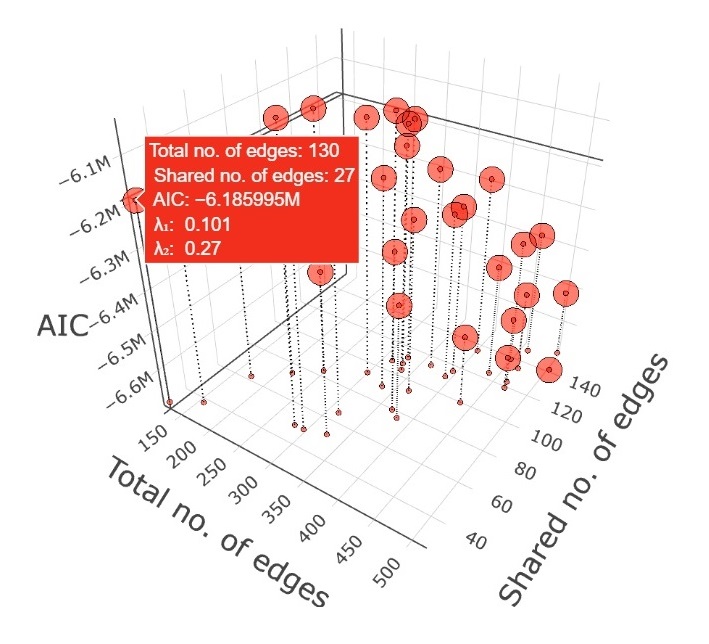}
  \caption{Interactive plot generated using the \texttt{pared} R package, displaying the set of all Pareto-optimal networks obtained by fitting the group graphical lasso to the gynecological cancer proteomic dataset. Hovering over any point reveals the corresponding tuning parameter values and the values of the multi-objective criteria at that solution.}
  \label{pared_interactive_plot}
\end{figure}

\section{Conclusion}
In this work, we introduced \texttt{pared}, a flexible and extensible R package for model selection through multi-objective optimization. Unlike traditional approaches that optimize a single criterion—such as cross-validation error or information criteria—\texttt{pared} enables identification of a set of Pareto-optimal models that balance competing objectives, including sparsity, interpretability, and structural similarity across groups. The package supports several widely used penalized models, including the elastic net, fused lasso, fused graphical lasso, and group graphical lasso, with model evaluation guided by context-appropriate metrics. Optimization is performed using Gaussian process-based surrogate modeling, allowing for efficient and scalable exploration of the hyperparameter space.

We demonstrated the utility of \texttt{pared} in a case study involving proteomics networks across gynecological cancers represented in The Cancer Proteome Atlas. The method identified a diverse set of high-quality models, some of which offered greater sparsity or structural coherence than those selected by traditional single-objective criteria such as AIC. An interactive visualization tool further enhances interpretability and supports user-guided model exploration.

Overall, \texttt{pared} provides a principled, reproducible, and interpretable framework for multi-objective model selection, with broad relevance to statistical learning and computational biology. Future work may extend the package to incorporate additional model classes and optimization strategies, further expanding its applicability.

\section*{Additional information}

\subsection*{Funding}
PD was partially supported by NIH/NCI Cancer Center Support Grant P30 CA016059. 
SR was partially supported by NSF Graduate Research Fellowship DGE 1842494 and  NIH T32 Grant CA96520-13: Training program in Biostatistics for Cancer Research.
CBP was partially supported by NIH/NCI R01 CA244845 and NIH/NCI CCSG P30CA016672 (Biostatistics Resource Group).

\vspace{-0.3cm}
\bibliographystyle{apalike}
\bibliography{reference}

\begin{thebibliography}{}

\bibitem[Akaike, 1974]{AIC}
Akaike, H. (1974).
\newblock A new look at the statistical model identification.
\newblock {\em IEEE T Automat Contr}, 19(6):716--723.

\bibitem[Berger et~al., 2018]{Berger2018}
Berger, A., A., K., et~al. (2018).
\newblock A comprehensive pan-cancer molecular study of gynecologic and breast cancers.
\newblock {\em Cancer Cell}, 33(4):690--705.

\bibitem[Collette and Siarry, 2003]{collette2003multiobjective}
Collette, Y. and Siarry, P. (2003).
\newblock {\em Multiobjective Optimization: Principles and Case Studies}.
\newblock Springer-Verlag, Berlin, Heidelberg.

\bibitem[Danaher et~al., 2014]{jgl}
Danaher, P., Wang, P., and Witten, D.~M. (2014).
\newblock The joint graphical lasso for inverse covariance estimation across multiple classes.
\newblock {\em J Roy Stat Soc B}, 76(2):373--397.

\bibitem[Das et~al., 2020]{Das2020}
Das, P. et~al. (2020).
\newblock {NExUS}: Bayesian simultaneous network estimation across unequal sample sizes.
\newblock {\em Bioinformatics}, 36(3):798--804.

\bibitem[Friedman et~al., 2008]{friedman2008sparse}
Friedman, J., Hastie, T., and Tibshirani, R. (2008).
\newblock Sparse inverse covariance estimation with the graphical lasso.
\newblock {\em Biostatistics}, 9(3):432--441.

\bibitem[Karl et~al., 2023]{karl2023multi}
Karl, F., Pielok, T., Moosbauer, J., Pfisterer, F., Coors, S., et~al. (2023).
\newblock Multi-objective hyperparameter optimization in machine learning—an overview.
\newblock {\em ACM Trans Evol Learn Optim}, 3(4):1--50.

\bibitem[Li et~al., 2013a]{Li2013TCPA}
Li, J. et~al. (2013a).
\newblock {TCPA}: a resource for cancer functional proteomics data.
\newblock {\em Nat Methods}, 10:1046--1047.

\bibitem[Li et~al., 2013b]{Li2013}
Li, S., Hsu, L., Peng, J., and Wang, P. (2013b).
\newblock Bootstrap inference for network construction with an application to a breast cancer microarray study.
\newblock {\em Ann Appl Stat}, 7(1):391.

\bibitem[Liu et~al., 2010]{stars}
Liu, H., Roeder, K., and Wasserman, L. (2010).
\newblock Stability approach to regularization selection {(StARS)} for high dimensional graphical models.
\newblock {\em Adv Neur In}, 23.

\bibitem[Meinshausen and B{\"u}hlmann, 2010]{stability_selection}
Meinshausen, N. and B{\"u}hlmann, P. (2010).
\newblock Stability selection.
\newblock {\em J Roy Stat Soc B}, 72(4):417--473.

\bibitem[Roberts and Nowak, 2014]{roberts2014stabilizing}
Roberts, S. and Nowak, G. (2014).
\newblock Stabilizing the lasso against cross-validation variability.
\newblock {\em Comput Stat Data An}, 70:198--211.

\bibitem[Schwarz, 1978]{BIC}
Schwarz, G. (1978).
\newblock Estimating the dimension of a model.
\newblock {\em Ann Stat}, pages 461--464.

\bibitem[Snoek et~al., 2012]{Snoek2012}
Snoek, J., Larochelle, H., and Adams, R.~P. (2012).
\newblock Practical {B}ayesian optimization of machine learning algorithms.
\newblock {\em Adv Neur In}, 25.

\bibitem[Tibshirani, 1996]{lasso}
Tibshirani, R. (1996).
\newblock Regression shrinkage and selection via the lasso.
\newblock {\em J Roy Stat Soc B}, 58(1):267--288.

\bibitem[Tibshirani et~al., 2005]{flasso}
Tibshirani, R., Saunders, M., Rosset, S., Zhu, J., and Knight, K. (2005).
\newblock Sparsity and smoothness via the fused lasso.
\newblock {\em J Roy Stat Soc B}, 67(1):91--108.

\bibitem[van Nee et~al., 2023]{van2023fast}
van Nee, M.~M., van~de Brug, T., and van~de Wiel, M.~A. (2023).
\newblock Fast marginal likelihood estimation of penalties for group-adaptive elastic net.
\newblock {\em J Comput Graph Stat}, 32(3):950--960.

\bibitem[Wu and Wang, 2020]{wu2020survey}
Wu, Y. and Wang, L. (2020).
\newblock A survey of tuning parameter selection for high-dimensional regression.
\newblock {\em Annu Rev Stat Appl}, 7(1):209--226.

\bibitem[Xue et~al., 2012]{xue2012particle}
Xue, B., Zhang, M., and Browne, W.~N. (2012).
\newblock Particle swarm optimization for feature selection in classification: A multi-objective approach.
\newblock {\em IEEE T Cybernetics}, 43(6):1656--1671.

\bibitem[Zou and Hastie, 2005]{enet}
Zou, H. and Hastie, T. (2005).
\newblock Regularization and variable selection via the elastic net.
\newblock {\em J Roy Stat Soc B}, 67(2):301--320.

\bibitem[Zou et~al., 2007]{Zou2007}
Zou, H., Hastie, T., and Tibshirani, R. (2007).
\newblock On the “degrees of freedom” of the lasso.
\newblock {\em Ann Stat}, 35(5):2173--2192.

\end{thebibliography}

\end{document}


\maketitle
\thispagestyle{empty}


\section*{A. Computation times}
In the accompanying GitHub vignette\footnote{\url{https://github.com/priyamdas2/pared}}, we present a set of example scenarios in which Pareto-optimal models are identified using the \texttt{pared} R package, fitting elastic net regression, fused lasso regression, and joint graphical lasso with both \textit{fused} and \textit{group} penalties. All computations are performed on a desktop running the Windows 10 Enterprise operating system desktop with 32 GB RAM and the following processor characteristics:  12th Gen Intel(R) Core(TM) i7-12700, 2100 Mhz, 12 Cores(s), 20 Logical Processor(s). The  computation times and other model parameter details are provided below in Table~\ref{comp_times}. The parameter \textit{Pareto Budget}, set by the user, denotes the maximum number of objective function evaluations allowed during the multi-objective optimization (MOO) process within \texttt{pared} functions.

\begin{table}[h]
\centering
\resizebox{0.7\columnwidth}{!}{%
\begin{tabular}{llcccc}
\hline
R functions & Model & Parameters & Sample size & Pareto Budget & \begin{tabular}[c]{@{}c@{}}Computation\\ time (sec.)\end{tabular} \\ \hline
\texttt{pared\_ENet} & Elastic net & $p=5$ & 100 & 50 & 95.05 \\
\texttt{pared\_FLasso} & Fused lasso & $p=10$ & 100 & 80 & 170.31 \\
\texttt{pared\_JGL} & Fused JGL & $K = 4, p = 20$ & (20, 40, 60, 80) & 40 & 166.91 \\
\texttt{pared\_JGL} & Group JGL & $K = 4, p = 20$ & (30, 50, 40, 70) & 50 & 148.54 \\ \hline
\end{tabular}}
\caption{Computation times (in seconds) required to identify the optimal set of Pareto-optimal models using the \texttt{pared} R package functions: \texttt{pared\_ENet}, \texttt{pared\_FLasso}, and \texttt{pared\_JGL}.
\label{comp_times}}
\end{table}
For the case study involving proteomic networks (each of size 20) for ovarian (OV), uterine corpus endometrial carcinoma (UCEC), and uterine carcinosarcoma (UCS) cancers—with sample sizes of 428, 404, and 48, respectively—it took 141 seconds (2.4 minutes) to estimate the Pareto-front curve using the \texttt{pared\_JGL} function with the \textit{group} penalty.
